\documentclass[letterpaper]{article} 
\usepackage{aaai2026}  
\usepackage{times}  
\usepackage{helvet}  
\usepackage{courier}  
\usepackage[hyphens]{url}  
\usepackage{graphicx} 
\usepackage{natbib}  
\usepackage{caption} 
\usepackage[dvipsnames]{xcolor}
\usepackage{booktabs}
\usepackage{mathrsfs}
\usepackage{xspace}
\usepackage{amsthm}
\usepackage{amsmath,amssymb,amsfonts}
\usepackage{algorithm}
\usepackage{algpseudocode}
\usepackage{array}
\usepackage{multirow}
\usepackage{fontawesome5}
\usepackage[caption=false,font=footnotesize,labelfont=rm,textfont=rm]{subfig}
\usepackage{makecell}
\usepackage{textcomp}
\usepackage{url}
\usepackage{verbatim}
\usepackage{graphicx}
\usepackage{cite}
\newcommand{\good}{\checkmark}
\newcommand{\medium}{--}
\newcommand{\bad}{\ding{55}}

\newcommand{\ourmethod}{\textsc{ARG-Designer}\xspace}

\usepackage{pifont}
\usepackage{newfloat}
\usepackage{listings}
\DeclareCaptionStyle{ruled}{labelfont=normalfont,labelsep=colon,strut=off} 
\floatstyle{ruled}
\newfloat{listing}{tb}{lst}{}
\floatname{listing}{Listing}
\title{Assemble Your Crew: Automatic Multi-agent Communication Topology Design via Autoregressive Graph Generation}
\author{
    Shiyuan Li\textsuperscript{\rm 1}, Yixin Liu\textsuperscript{\rm 1}, Qingsong Wen\textsuperscript{\rm 2}, Chengqi Zhang\textsuperscript{\rm 3}, Shirui Pan\textsuperscript{\rm 1}\thanks{Corresponding Author}
}

\affiliations{
    \textsuperscript{\rm 1}Griffith University\\
    \textsuperscript{\rm 2}Squirrel Ai Learning\\
    \textsuperscript{\rm 3}Hong Kong Polytechnic University\\
    \{li.shiy511, qingsonedu\}@gmail.com, \{yixin.liu, s.pan\}@griffith.edu.au, chengqi.zhang@polyu.edu.hk
}

\begin{document}

\maketitle

\begin{abstract}
Multi-agent systems (MAS) based on large language models (LLMs) have emerged as a powerful solution for dealing with complex problems across diverse domains. The effectiveness of MAS is critically dependent on its collaboration topology, which has become a focal point for automated design research. However, existing approaches are fundamentally constrained by their reliance on a template graph modification paradigm with a predefined set of agents and hard-coded interaction structures, significantly limiting their adaptability to task-specific requirements. To address these limitations, we reframe MAS design as a conditional autoregressive graph generation task, where both the system composition and structure are designed jointly. We propose \textbf{\ourmethod}, a novel autoregressive model that operationalizes this paradigm by constructing the collaboration graph from scratch. Conditioned on a natural language task query, \ourmethod sequentially and dynamically determines the required number of agents, selects their appropriate roles from an extensible pool, and establishes the optimal communication links between them. This generative approach creates a customized topology in a flexible and extensible manner, precisely tailored to the unique demands of different tasks. Extensive experiments across six diverse benchmarks demonstrate that \ourmethod not only achieves state-of-the-art performance but also enjoys significantly greater token efficiency and enhanced extensibility.
\end{abstract}

\begin{links}
    \link{Code}{https://github.com/Shiy-Li/ARG-Designer}
\end{links}

\section{Introduction}
Agents built on large language models (LLMs) have demonstrated impressive capabilities in tackling complex tasks across domains, including code generation, data analysis, decision-making, and question answering~\cite{zhu2024large,li2024dawn,song2023llm,wang2024chain,zhong2024debug,tan2025bisecle,miao2025blindguard}. To overcome the limitations of a single agent in tackling more complex tasks, the research interests have increasingly shifted towards multi-agent systems (MAS), which unlock new potential through collaborative interactions among agents with diverse capabilities and roles. Central to MAS is its \textbf{collaboration topology}, a graph that defines how agents with roles are structured and how they exchange information~\cite{liu2025graph}. A growing body of evidence, spanning from sequential reasoning pipelines to debate-based approaches, demonstrates that MAS performance varies dramatically depending on how inter-agent communication is architected~\cite{zhang2024aflow,zhou2025multi}. Therefore, designing an effective collaboration graph tailored to specific tasks becomes a critical research challenge. 

\begin{figure}[t]
    \centering
    \includegraphics[width=\linewidth]{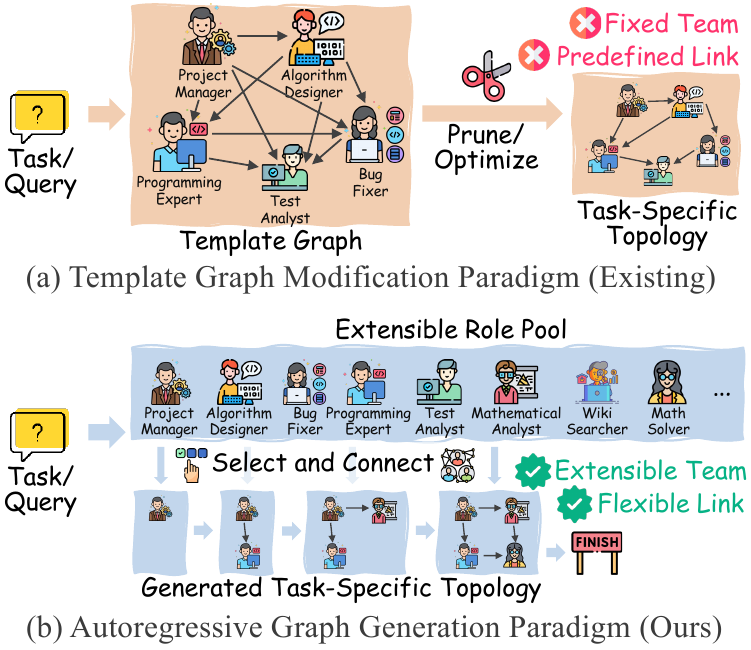}
    \caption{The comparison of two paradigms: (a)~template graph modification and (b)~autoregressive graph generation.}
    \label{fig:concept}
\end{figure}

Early research on MAS topology design focused on static and manually designed graphs, such as chains that enforce sequential workflow~\cite{wei2022chain,hong2023metagpt}, trees that enable structured deliberation~\cite{yao2023tree}, and fully connected graphs that ensure sufficient communication. Although these canonical collaboration topologies can facilitate effective coordination in specific scenarios, the inherent rigidity of these fixed topologies limits their adaptability across diverse tasks, resulting in sub-optimal performance. To enhance flexibility and efficiency, a more recent line of work~\cite{zhuge2024gptswarm,shen2025understanding} focuses on adaptively constructing task-specific communication structures using \textit{graph learning models}~\cite{pan2025survey,pan2025label,fu2025less,fu2025learn}. For example, AgentPrune~\cite{zhang2024cut} and AgentDrop~\cite{wang2025agentdropout} learn to create sparse and task-specific graphs by pruning connections or agents from a predefined template topology.
More advanced approaches like G-Designer~\cite{zhang2024g} follow the paradigm of graph structure learning, which uses a graph autoencoder to learn efficient collaboration structures in a task-adaptive manner.

Despite their varied designs, existing graph learning-based methods often follow a shared paradigm: \textit{template graph modification} (Fig.~\ref{fig:concept}a). That is, they typically start from a fixed communication template based on a predefined set of agents and hard-coded interaction structures, and apply learnable adjustments, such as edge reweighting or pruning, to adapt the topology to specific tasks~\cite{zhang2024cut,zhang2024g}. Despite offering reasonable adaptability in constrained settings, this paradigm exhibits two inherent limitations. \textbf{\textit{Limitation~1: Redundant Composition}}. To ensure structural flexibility, template graphs are often initialized with numerous agent roles and densely connected edges, many of which are unnecessary for a specific task. Even with pruning mechanisms, irrelevant agents or connections may be retained in the learned task-specific topology, leading not only to reduced efficiency but also to potential sub-optimal decision-making during execution. \textbf{\textit{Limitation~2: Limited Extensibility}}. In the fast-evolving field of LLM-based agents, a massive number of new agent functionalities are emerging with increasing frequency. However, trained on a fixed template graph, the existing methods struggle to generalize to scenarios with dynamic agent sets or evolving collaboration needs. Meanwhile, it would be prohibitively expensive to build a large-scale template graph that covers all possible agent roles and interaction patterns, and then prune it to a suitable task-specific topology. Given the above limitations, a natural question arises: \textit{Going beyond template graph modification, can we design a more flexible and extensible paradigm for collaboration topology construction?}

To seek the answer to the above question, we draw inspiration from real-world practices of recruiting teams for complex tasks. Rather than starting with a fully staffed team where every possible member is onboarded from the beginning, real-world teams are usually formed incrementally, with members added based on expertise, availability, and evolving task needs. This practical pattern inspires us to explore \textit{\textbf{autoregressive graph generation}} (Fig.~\ref{fig:concept}b) as a more promising paradigm for collaboration topology construction. Unlike pruning from a predefined overcomplete structure, the new generation paradigm constructs the collaboration graph from scratch by progressively selecting appropriate agents. Such an incremental procedure naturally avoids redundant agent-role compositions during the design process, which naturally addresses \textit{\textbf{Limitation~1}}. Moreover, by discarding the fixed template, the generative paradigm enables dynamic expansion of the agent pool, with only linear computational cost during the node generation phase. This merit enhances the extensibility of collaboration graph construction and thus alleviates \textit{\textbf{Limitation~2}}.

Building upon the new paradigm, in this paper, we propose \textbf{\ourmethod}, a novel \textbf{{A}}uto\textbf{{R}}egressive \textbf{{G}}raph generation model that acts as a MAS topology \textbf{{Designer}}. Conditioned on a natural language task query, \ourmethod constructs the entire collaboration graph from scratch by iteratively generating each node (i.e., agent) along with its corresponding edges (i.e., communication links) to previously generated nodes. Compared to prior approaches, \ourmethod provides enhanced flexibility and scalability with respect to the number of agents, the variety of agent roles, and the richness of potential interactions. To train our generative model, we design a curriculum learning strategy that starts with denser communication topologies to ease the cold-start problem, and gradually transitions to sparser, pruned graphs for fine-tuning, encouraging the model to generalize to minimal yet effective structures. Extensive experiments on six benchmarks demonstrate that our method achieves state-of-the-art effectiveness, communication efficiency, and robustness.

\section{Problem Formulation}\label{sec:problem_formulation}
In this section, we introduce the graph-based modeling of MAS, and then formulate MAS topology design as an autoregressive graph generation problem.

\noindent\textbf{MAS as a Collaboration Graph.} 
We model a MAS as a \textit{collaboration graph}, a directed acyclic graph (DAG) $\mathcal{G} = (\mathcal{V}, \mathcal{E})$ that outlines the system architecture and the flow of information among its components. The nodes $\mathcal{V} = \{v_1, v_2, \dots, v_N\}$ represent the set of agents, where each agent $v_i$ is an instance of an LLM endowed with a specific role $r_i \in R$ that dictates its function and expertise. It also maintains an internal state $s_i \in S$, which serves as a memory of its past actions and interactions. The edges $\mathcal{E} \subseteq \mathcal{V} \times \mathcal{V}$ define the directed communication pathways. An edge $e_{ji} = (v_j, v_i)$ signifies that agent $v_i$ is a designated recipient of information from agent $v_j$. The set of direct predecessors of agent $v_i$ is denoted by ${\mathcal{N}}_{in}(v_i) = \{v_j \mid (v_j, v_i) \in \mathcal{E}\}$.

\noindent\textbf{MAS Collaboration Protocol.} 
Given a collaboration graph $\mathcal{G}$, the MAS addresses a user query $\mathcal{Q}$ by executing a multi-step collaboration protocol. This protocol governs how information is processed and passed between agents, unfolding over a series of communication rounds~\cite{pan2026correcting,zhang2024trustworthy,cai2024fgad}. 
Unlike traditional GNNs message passing, the operational sequence for agent activation within each round is determined by a topological sort of the nodes, ensuring that an agent is activated only after its prerequisite inputs are available~\cite{zhao2025freegad,chen2025uncertainty,zhuang2025refine,li2024noise,cai2025agdiff}. This process can be executed for $K$ rounds to allow for iterative refinement. In each round $k \in \{1, \dots, K\}$, an agent $v_i$ generates its response $m_i^{(k)}$ by invoking its language model with a dynamically constructed prompt $\mathcal{P}_i^{(k)}$:
\begin{equation}
    m_i^{(k)} = \text{LLM}_i(\mathcal{P}_i^{(k)}).
\end{equation}
\noindent where the prompt integrates the properties of the agent with outputs of its predecessors from the previous round:
\begin{equation}
    \mathcal{P}_i^{(k)} = f(\underbrace{r_i, s_i}_{\texttt{System}}, \underbrace{\mathcal{Q}, \{m_j^{(k-1)} \mid v_j \in \mathcal{N}_{in}(v_i)\}}_\texttt{User}).
\end{equation}

After $K$ rounds, the final output $\mathcal{O}$ is obtained by aggregating the final-round responses from some or all agents:
\begin{equation}
    \mathcal{O} = \text{Aggregate}(\{m_i^{(K)} \mid v_i \in \mathcal{V} \}),
\end{equation}
where the aggregation strategy $\text{Aggregate}(\cdot)$ varies across implementations. Common strategies include majority voting, delegating the final decision to a specific terminal agent, or selecting the output from the last agent in the execution order. The number of communication rounds $K$ can be either predefined or adaptively determined via early-stopping.

\noindent\textbf{MAS Topology Design as a Graph Generation Task.} 
The automatic task-specific construction of MAS topologies is a key challenge and research frontier. Traditional automated approaches that start from a large, predefined template graph, analogous to a fully-staffed team with every possible role, suffer from redundancy and limited extensibility. Drawing inspiration from real-world practices of building expert teams incrementally, we reframe the problem from modifying a fixed template to generating a bespoke graph from scratch. Instead of navigating the enormous graph space $\mathbb{G}$ with an expensive utility function $\phi(\text{Execute}(\mathcal{G}, \mathcal{Q}))$, we propose to learn a \textbf{conditional generative model}, i.e.,  $P(\mathcal{G} | \mathcal{Q}, \mathcal{R})$, where $\mathcal{R}$ is an extensive agent role pool. This model directly captures the relationship between a task query and the principles of effective collaboration, aiming to find the optimal communication topology $\mathcal{G}^*$ that is most probable under this learned distribution:
\begin{equation}
\label{eq:gen_formulation}
\mathcal{G}^* = \arg\max_{\mathcal{G} \in \mathbb{G}} P(\mathcal{G} | \mathcal{Q}, \mathcal{R}).
\end{equation}

Compared to modifying a predefined template graph, the generative formulation offers a more flexible, extensible, and scalable approach for constructing high-quality MAS topologies.

\noindent\textbf{Autoregressive Graph Generation.} 
To make this topology generation process more tractable, we formulate it as an autoregressive graph generation problem. This formulation decomposes the intractable joint probability of an entire graph into a tractable sequence of conditional probabilities. The graph is constructed incrementally, where each step involves adding a new node and its corresponding edges, conditioned on the partial graph built so far.

Formally, this factorization is expressed as:
\begin{equation}
\label{eq:autoregressive_formulation}
P(\mathcal{G} | \mathcal{Q}, \mathcal{R}) = \prod_{i=1}^{|\mathcal{V}|} \underbrace{P(v_i | \mathcal{G}_{<i}, \mathcal{Q},\mathcal{R})}_{\text{Node Generation}} \cdot \underbrace{\prod_{j=1}^{i-1} P(e_{ji} | v_i, \mathcal{G}_{<i}, \mathcal{Q})}_{\text{Edge Generation}},
\end{equation}
where $\mathcal{G}_{<i}$ represents the subgraph of the first $i-1$ nodes. The generation process at each step $i$ thus involves two key actions: \textbf{node generation}, predicting the role of the next agent to add, and \textbf{edge generation}, establishing its connections from existing agents. This formulation provides significant flexibility, enabling the model to dynamically determine the total number of agents by learning to sample a special \texttt{END} token, and to model complex structural dependencies by conditioning on the generation history.

\noindent\textbf{Discussion.} 
This generative approach provides several key advantages over traditional template-based methods, as summarized in Table~\ref{tab:freedom_comparison}. \ding{182}~\textbf{Task-Adaptive Construction.} By conditioning on the task query, the model constructs a bespoke collaboration graph from scratch, avoiding the rigidity and one-size-fits-all limitations of predefined graphs. \ding{183}~\textbf{Dynamic and Extensible Composition.} The model dynamically determines the necessary number of agents and selects their roles from an extensible pool, ensuring the MAS composition is precisely tailored to the task needs and can easily incorporate new agent capabilities. \ding{184}~\textbf{Tractable Generation.} The autoregressive factorization transforms the intractable problem of generating a whole graph into a sequence of simple, conditional steps, making the learning process both manageable and scalable.

\begin{table}[t!]
\centering
\small
\begin{tabular}{lccc}
\toprule
\multirow{2}{*}{\textbf{Method}} & {\textbf{Task-}} & {\textbf{Variable}} & {\textbf{Flexible}} \\
 &\textbf{Adaptive}&\textbf{Size}& \textbf{Roles} \\
\midrule
\textbf{Manual Design}    & \bad      & \bad      & \bad \\
\midrule
\textbf{AgentDropout}     & \good      & \bad      & \medium \\
\textbf{AgentPrune}       & \good      & \medium      & \medium \\
\textbf{G-Designer}       & \good     & \bad      & \medium  \\
\midrule
\textbf{\ourmethod} (ours)       & \good     & \good     & \good \\
\bottomrule
\end{tabular}
\caption{Comparison of degrees of freedom in MAS design paradigms. The icons {\good}, {\medium}, and {\bad} represent full, partial, and no support for each capability, respectively.}
\label{tab:freedom_comparison}
\end{table}

\section{\ourmethod for MAS Topology Design}
\begin{figure*}[t]
    \centering
    \includegraphics[width=1.0\linewidth]{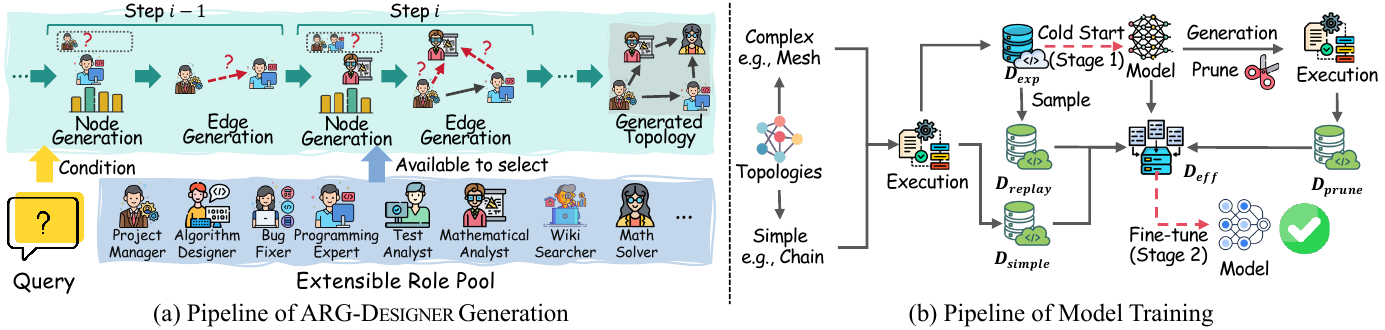}
    \vspace{-3mm}
    \caption{The pipeline of \ourmethod, including (a) MAS communication topology generation and (b) model training.}
    \label{fig:pipeline}
\end{figure*}
Based on the autoregressive graph generation paradigm, this section instantiates the proposed method, \textbf{\ourmethod}, which is specifically crafted for MAS topology generation. We first introduce the model architecture designed to implement the sequential generation process, and then describe the training and inference strategies to guide \ourmethod toward generating both functionally correct and structurally efficient collaboration graphs.

\subsection{Model Architecture}
Following the autoregressive generation paradigm, {\ourmethod} constructs collaboration graphs step-by-step. \ourmethod employs a hierarchical architecture based on gated recurrent units (GRUs), which are well-suited for sequence modeling due to their effectiveness in capturing long-range dependencies while maintaining computational efficiency. The architecture separates the generation model into two sub-components: a \textit{node generator} to select agent roles and an \textit{edge generator} to build communication links. An overview of the model architecture is depicted in Fig.~\ref{fig:pipeline}a.

\noindent\textbf{Input Representation.} 
Before generation begins, \ourmethod encodes all textual conditioning information (i.e., the task query and available agent roles) into dense vector representations. Specifically, the task query $\mathcal{Q}$ is mapped into a fixed-dimensional vector $\mathbf{f}_\mathcal{Q}\in \mathbb{R}^{d}$ by a pre-trained BERT-like sentence encoder followed by a feed-forward network (FFN) with Layer Normalization (LN):
\begin{equation}
\mathbf{f}_\mathcal{Q} = \text{FFN}(\text{LN}(\text{SentenceEncoder}(\mathcal{Q}))).
\end{equation}

Similarly, each available agent role $r_k \in \mathcal{R}$ is converted into an embedding $\mathbf{z}_{r_k}$. These pre-computed embeddings are collected into a role embedding matrix $\mathbf{Z} \in \mathbb{R}^{|\mathcal{R}| \times d}$, serving as the knowledge base of available agents.

\noindent\textbf{Node Generation.} 
At each step $i$, the node generator selects the role for the next agent, i.e., $v_i$. 
\ourmethod first models the context information by combining task information with the generation history. A dedicated GRU, $\text{GRU}_{\text{prev}}$, is employed to aggregate the role embeddings of all preceding agents to form a historical embedding $\mathbf{f}^{(i)}_{\text{hist}}$:
\begin{equation}
    \mathbf{f}^{(i)}_{\text{hist}} = \text{GRU}_{\text{prev}}([\mathbf{z}_{r_1}, \mathbf{z}_{r_2}, \ldots, \mathbf{z}_{r_{i-1}}]).
\end{equation}

Then, we fuse the historical embedding $\mathbf{f}_{\text{hist}}^{(i)}$ with the task embedding $\mathbf{f}_\mathcal{Q}$ via a dynamic gate to produce the context embedding $\mathbf{f}_{\text{cont}}^{(i)}$:
\begin{equation}
    \mathbf{f}^{(i)}_{\text{cont}} = (1-\mathbf{g}_{i}) \cdot \mathbf{f}^{(i)}_{\text{hist}} + \mathbf{g}_{i} \cdot \mathbf{f}_\mathcal{Q}, \quad \mathbf{g}_{i} = \sigma\left(\frac{\mathbf{f}^{(i)}_{\text{hist}} \cdot \mathbf{f}_\mathcal{Q}}{\sqrt{d}}\right),
\end{equation}
where $\sigma$ denotes the sigmoid function. This context, along with an edge feature vector $\mathbf{f}^{(i)}_{\text{edge}}$ (i.e., a vector encoding the connectivity pattern of the previously added node $v_{i-1}$), are concatenated into an input vector $[\mathbf{f}^{(i)}_{\text{cont}}, \mathbf{f}^{(i)}_{\text{edge}}]$. Then, another GRU module, $\text{GRU}_{\text{node}}$, updates its hidden state $\mathbf{h}^{(i)}_{\text{node}}$, which captures the full generation condition:
\begin{equation}
    \mathbf{h}^{(i)}_{\text{node}} = \text{GRU}_{\text{node}}(\text{MLP}_{\text{node}}([\mathbf{f}^{(i)}_{\text{cont}}, \mathbf{f}^{(i)}_{\text{edge}}]), \mathbf{h}^{(i-1)}_{\text{node}}).
\end{equation}

To preserve the extensible property of \ourmethod during agent role selection, we use a metric learning-based module for node generation. Concretely, the hidden state $\mathbf{h}^{(i)}_{\text{node}}$ is projected into a ``node intent'' embedding. Then, the node prediction scores $\mathbf{s}^{(i)}_{\text{node}}$ will be acquired by a dot-product operation with projected role embeddings:

\begin{equation}
    \mathbf{s}^{(i)}_{\text{node}} = \text{MLP}_{\text{pred\_n}}(\mathbf{h}^{(i)}_{\text{node}}) \cdot \text{MLP}_{\text{role}}([\mathbf{Z},\mathbf{z}_{\text{end}}]),
\end{equation}
\noindent where, $\mathbf{z}_{\text{end}}$ is a learnable embedding for ending token \texttt{END}, which signals the termination of the generation process. 
Finally, we can obtain the predicted probability as follows:
\begin{equation}
    P(v_i | \mathcal{G}_{<i}, \mathcal{Q},\mathcal{R}) = \text{Softmax}(\mathbf{s}^{(i)}_{\text{node}}),
\end{equation}
where the $\text{Softmax}(\cdot)$ function converts the scores into a probability distribution. 

\noindent\textbf{Discussion of Extensibility:} The design of the node generator in \ourmethod allows new agent roles to be added at inference time without retraining. When new roles are introduced, we can extend the role embedding matrix $\mathbf{Z}$ by appending new role-specific embedding rows. Since $\mathbf{s}^{(i)}_{\text{node}}$ is produced by a metric learning-based retrieval mechanism rather than a fixed-dimensional classifier, the model can flexibly select from an expanded set of roles based on similarity in the shared embedding space. This design ensures that \ourmethod remains extensible and adaptable to evolving agent pools, which well fits the real-world scenarios where new agents with novel functionalities are frequently introduced to meet emerging task demands.

\noindent\textbf{Edge Generation.} 
Once agent node $v_i$ is chosen, the edge generator determines its incoming connections from existing agents $\{v_1, \dots, v_{i-1}\}$. 
Here, we use a dedicated GRU, $\text{GRU}_{\text{edge}}$ to model this sequential process. Its hidden state is initialized from the final state of the node-level GRU, $\mathbf{h}^{(i)}_{\text{node}}$, serving as the condition of edge prediction:
\begin{equation}
    \mathbf{h}^{(i,0)}_{\text{edge}} = \text{MLP}_{\text{node2edge}}(\mathbf{h}^{(i)}_{\text{node}}).
\end{equation}

After that, the model iterates through previously predicted nodes $v_j$ ($j=1,\cdots,i-1$). At each sub-step, the edge GRU updates its hidden state based on the embedding of the previous edge decision:
\begin{equation}
    \mathbf{h}^{(i,j)}_{\text{edge}} = \text{GRU}_{\text{edge}}(\text{MLP}_{\text{edge}}(\mathbf{e}^{(j-1,i)}), \mathbf{h}^{(i,j-1)}_{\text{edge}}),
\end{equation}
where $\mathbf{e}^{(j-1,i)}$ is a one-hot vector representing the previous decision on whether to form an edge from node $v_{j-1}$ to $v_i$. Following that, the updated state $\mathbf{h}^{(i,j)}_{\text{edge}}$ is passed through an output MLP to predict the score:
\begin{equation}
    {s}^{(i,j)}_{\text{edge}} = \text{MLP}_{\text{pred\_e}}(\mathbf{h}^{(i,j)}_{\text{edge}}).
\end{equation}

We can then obtain the probability of edge $e_{j, i}$ by:
\begin{equation}
    P(e_{j,i}=1 | v_i, \mathcal{G}_{<i}, \mathcal{Q}) = \text{Sigmoid}({s}^{(i,j)}_{\text{edge}}).
\end{equation}

\subsection{Training and Inference Strategy}
\begin{algorithm}[t!]
\caption{The Inference algorithm  of \ourmethod}
\label{alg:inference}
\begin{algorithmic}[1]
\State \textbf{Input:} Task query $\mathcal{Q}$, trained model $P_\theta$
\State \textbf{Output:} Collaboration graph $\mathcal{G} = (\mathcal{V}, \mathcal{E})$
\State Initialize $\mathcal{V} \leftarrow \emptyset$, $\mathcal{E} \leftarrow \emptyset$, $i \leftarrow 1$
\Loop
    \State Sample agent role $r_i \sim P_\theta(v_i | \mathcal{G}_{<i}, \mathcal{Q})$
    \If{$r_i = \texttt{END}$ or $i > N_{\max}$}
        \State \textbf{break}
    \EndIf
    \State $\mathcal{V} \leftarrow \mathcal{V} \cup \{v_i\}$
    \For{$j = i-1$ \textbf{down to} $1$}
        \State Sample edge existence $b_{ji} \sim P_\theta(e_{ji} | v_i, \mathcal{G}_{<i}, \mathcal{Q})$
        \If{$b_{ji} = 1$}
            \State $\mathcal{E} \leftarrow \mathcal{E} \cup \{e_{ji}\}$
        \EndIf
    \EndFor
    \State $i \leftarrow i + 1$
\EndLoop
\State \Return $\mathcal{G} = (\mathcal{V}, \mathcal{E})$
\end{algorithmic}
\end{algorithm}
\subsubsection{Data Construction of Curriculum Learning.}

To build a powerful graph generator for MAS communication topology, we set up two key objectives: 
\ding{182}~\textbf{Functional correctness}, which ensures that the generated topology enables agents to collaboratively complete the given task, with all necessary roles and interactions properly instantiated; 
\ding{183}~\textbf{Communicational efficiency}, which encourages the generated topology to be lightweight, sparse, and compact, by minimizing redundant links or agents. To reach these goals, we design a \textbf{curriculum learning} strategy that constructs training data for two-stage training. In the first stage, we create an \textit{exploration dataset} to teach the model to generate correct and diverse topologies; then, an \textit{efficiency dataset} is built to guide the model to design simple yet communication-efficient topologies. 

Formally, we define a dataset as $\mathcal{D} = \{ (\mathcal{G}_k, \mathcal{Q}_k) \}_{k=1}^{M}$ that provides strong supervision on what constitutes an effective collaboration graph $\mathcal{G}_k$ for a given task query $\mathcal{Q}_k$. Since manually authoring such optimal task-graph pairs is infeasible, we propose to construct high-quality datasets in an automatically synthetic manner.

The first phase focuses on creating an {exploration dataset ($\mathcal{D}_{\text{exp}}$)} for the cold start training of the model, which aims to teach the model to create effective communication graphs. This dataset is formed by pairing tasks from a base set $\mathbb{Q}_{\text{base}}$ with resource-rich, complex configurations from a configuration space $\mathbb{C}_{\text{complex}}$, and retaining only empirically successful instances. Formally, this process is defined as:
\begin{equation}
    \mathcal{D}_{\text{exp}} = \{ (G(c), \mathcal{Q}) \mid , S(\mathcal{Q}, G(c)) = 1 \},
\end{equation}
where $\mathcal{Q} \in \mathbb{Q}_{\text{base}}$, $c \in \mathbb{C}_{\text{complex}}$ is a configuration blueprint specifying the high-level attributes of a graph, such as topology and agent count. For example, a configuration could be defined as $c = (\text{`star'}, 6, \mathcal{R})$. Note that the ``agent\_num'' parameter here does not limit the core capability of \ourmethod; rather, it constrains the data generation process to ensure a rich diversity of graph structures in the training data, from which the model learns generalizable collaborative patterns, not fixed sizes. $G(\cdot)$ is a deterministic function that maps a configuration $c$ to a specific graph instance $\mathcal{G}$, and $S(\mathcal{Q}, \mathcal{G}) \in \{0, 1\}$ is an indicator function that verifies the empirical success of the graph for the given task. This initial phase allows the model to learn fundamental collaborative patterns in an unconstrained and resource-abundant environment, which ensures the model has generalizable graph construction abilities.

In the second phase, an efficiency dataset $\mathcal{D}_{\text{eff}}$ is built to teach the model to generate more economical graphs. $\mathcal{D}_{\text{eff}}$ is a heterogeneous mixture composed of three sources:
\begin{equation}
    \mathcal{D}_{\text{eff}} = \mathcal{D}_{\text{simple}} \cup \mathcal{D}_{\text{pruned}} \cup \mathcal{D}_{\text{replay}},
\end{equation}
$\mathcal{D}_{\text{eff}}$ includes natively efficient graphs from simple configurations $\mathcal{D}_{\text{simple}}$, successful graphs derived from pruning the dense structures generated by the Phase 1 model $\mathcal{D}_{\text{pruned}}$, and a subset 
of the original exploration data $\mathcal{D}_{\text{replay}}$ to prevent catastrophic forgetting.
More specifically: \ding{182}~$\mathcal{D}_{\text{simple}}$, which contains task-graph pairs generated from a predefined set of minimal, manually-verified configurations known to be efficient; \ding{183}~$\mathcal{D}_{\text{pruned}}$, created by taking the overly complex but functional graphs from $\mathcal{D}_{\text{exp}}$, systematically removing individual nodes or edges, and retaining any pruned versions that still successfully complete the task; \ding{184}~$\mathcal{D}_{\text{replay}}$, a random subset of the initial $\mathcal{D}_{\text{exp}}$ dataset, included to prevent the model from forgetting the fundamental patterns learned in the first phase. With the carefully designed dataset for the second training stage, \ourmethod learns to strike a desirable balance between correctness and simplicity, producing high-quality topologies with minimal redundancy.

\subsubsection{Model Training.}
The training objective is to maximize the conditional log-likelihood of the ground-truth graphs in a given dataset $\mathcal{D}$. The model parameters $\theta$ are optimized by minimizing the negative log-likelihood (NLL) loss:
\begin{equation}
    \mathcal{L}(\theta) = - \sum_{(\mathcal{G}, \mathcal{Q}) \in \mathcal{D}} \log P_\theta(\mathcal{G} | \mathcal{Q}).
\end{equation}

Following the autoregressive factorization, the above loss is decomposed into a node generation term and an edge generation term. The final training loss is a weighted sum of these two terms:
\begin{equation}
    \mathcal{L}_{\text{total}} = \alpha \cdot \mathcal{L}_{\text{node}} + (1-\alpha) \cdot \mathcal{L}_{\text{edge}},
\end{equation}
where $\alpha \in [0, 1]$ is a hyperparameter balancing the two objectives. The individual loss terms are defined as the NLL over all node and edge generation steps, respectively:
\begin{align}
    \mathcal{L}_{\text{node}} &= - \sum_{(\mathcal{G}, \mathcal{Q}) \in \mathcal{D}} \sum_{i=1}^{|\mathcal{V}|} \log P_\theta(v_i | \mathcal{G}_{<i}, \mathcal{Q}, \mathcal{R}), \\
    \mathcal{L}_{\text{edge}} &= - \sum_{(\mathcal{G}, \mathcal{Q}) \in \mathcal{D}} \sum_{i=1}^{|\mathcal{V}|} \sum_{j=1}^{i-1} \log P_\theta(e_{ji} | v_i, \mathcal{G}_{<i}, \mathcal{Q}).
\end{align}

Throughout training, we employ a {teacher forcing} strategy, feeding the model ground-truth structures at each step to stabilize and accelerate learning. We train \ourmethod following a two-phase process, see Fig.~\ref{fig:pipeline}b. We begin with a \textbf{cold start} on $\mathcal{D}_{\text{exp}}$, followed by \textbf{efficiency fine-tuning} on $\mathcal{D}_{\text{eff}}$ with a lower learning rate.

\subsubsection{Inference.}
During inference, given a new task query $\mathcal{Q}$, the trained \ourmethod model generates a collaboration graph autoregressively without any ground-truth guidance. The process begins by initializing an empty graph $\mathcal{G} = (\mathcal{V}, \mathcal{E})$. It then enters a generation loop that iteratively builds the graph node by node. At each step $i$, the model's node generator first samples a role $r_i$ for the new agent $v_i$ from the probability distribution $P_\theta(v_i | \mathcal{G}_{<i}, \mathcal{Q})$. If the sampled role is the special \texttt{END} token, or if a predefined maximum number of agents $N_{\max}$ is reached, the generation process terminates. Otherwise, the new node $v_i$ is added to the vertex set $\mathcal{V}$. Subsequently, the model's edge generator is invoked. It sequentially considers each existing node $v_j \in \mathcal{G}_{<i}$ and samples the existence of an incoming edge $e_{ji}$ from $v_j$ to $v_i$. This entire process of node and edge generation is repeated until a termination condition is met, at which point the final graph $\mathcal{G}$ is returned. A detailed algorithmic description see Algo.~\ref{alg:inference}.

\begin{table}[t]
\resizebox{\columnwidth}{!}{
\centering
\begin{tabular}{llllll}
\toprule
\textbf{Category} & \textbf{Dataset} & \textbf{Answer Type} & \textbf{Metric} & \textbf{\#Test} & \textbf{License} \\
\midrule
General reasoning & MMLU & Multi-choice & Acc. & 153 & MIT License \\
\midrule
\multirow{4}{*}{Math reasoning} & GSM8K & Number & Acc. & 1,319 & MIT License \\
& MultiArith & Number & Acc. & 600 & Unspecified \\
& SVAMP & Number & Acc. & 1,000 & MIT License \\
& AQuA & Multi-choice & Acc. & 254 & Apache-2.0 \\
\midrule
Code generation & HumanEval & Code & Pass@1 & 164 & MIT License \\
\bottomrule
\end{tabular}
}
\caption{Dataset descriptions and statistics.}
\label{tab:dataset_statistics}
\end{table}

\begin{table*}[t]
\centering
\small
\begin{tabular}{l|p{1.7cm}<{\centering}p{1.7cm}<{\centering}p{1.7cm}<{\centering}p{1.7cm}<{\centering}p{1.7cm}<{\centering}p{1.7cm}<{\centering}|p{1.7cm}<{\centering}}
\toprule
\textbf{Method} & \textbf{MMLU} & \textbf{GSM8K} & \textbf{AQuA} & \textbf{MultiArith} & \textbf{SVAMP} & \textbf{HumanEval} & \textbf{Average} \\
\midrule
Vanilla                                & 80.39      & 82.30          & 71.06              & 93.09        & 86.55          & 71.39             & 80.80         \\ \midrule
CoT                              & 81.69 \scriptsize$\uparrow$1.30 &
86.50 \scriptsize$\uparrow$4.20 &
73.58 \scriptsize$\uparrow$2.52 &
93.25 \scriptsize$\uparrow$0.16 &
87.36 \scriptsize$\uparrow$0.81 &
74.67 \scriptsize$\uparrow$3.28 &
82.84 \scriptsize$\uparrow$2.04          \\
SC (CoT)               &
83.66 \scriptsize$\uparrow$3.27 &
81.60 \scriptsize$\downarrow$0.70 &
75.63 \scriptsize$\uparrow$4.57 &
94.12 \scriptsize$\uparrow$1.03 &
88.59 \scriptsize$\uparrow$2.04 &
79.83 \scriptsize$\uparrow$8.44 &
83.91 \scriptsize$\uparrow$3.11          \\
 \midrule
 Chain                              &83.01 \scriptsize$\uparrow$2.62 &
88.30 \scriptsize$\uparrow$6.00 &
74.05 \scriptsize$\uparrow$2.99 &
93.27 \scriptsize$\uparrow$0.18 &
87.17 \scriptsize$\uparrow$0.62 &
81.37 \scriptsize$\uparrow$9.98 &
84.53 \scriptsize$\uparrow$3.73          \\
Tree              &
81.04 \scriptsize$\uparrow$0.65 &
85.20 \scriptsize$\uparrow$2.90 &
71.23 \scriptsize$\uparrow$0.17 &
93.68 \scriptsize$\uparrow$0.59 &
88.91 \scriptsize$\uparrow$2.36 &
80.53 \scriptsize$\uparrow$9.14 &
83.43 \scriptsize$\uparrow$2.63        \\
Complete                           &
82.35 \scriptsize$\uparrow$1.96 &
80.10 \scriptsize$\downarrow$2.20 &
72.95 \scriptsize$\uparrow$1.89 &
94.53 \scriptsize$\uparrow$1.44 &
84.01 \scriptsize$\downarrow$2.54 &
79.03 \scriptsize$\uparrow$7.64 &
82.16 \scriptsize$\uparrow$1.36         \\
Random               &
84.31 \scriptsize$\uparrow$3.92 &
86.90 \scriptsize$\uparrow$4.60 &
76.48 \scriptsize$\uparrow$5.42 &
94.08 \scriptsize$\uparrow$0.99 &
87.54 \scriptsize$\uparrow$0.99 &
82.66 \scriptsize$\uparrow$11.27 &
85.33 \scriptsize$\uparrow$4.53        \\
LLM-Debate                           &
84.96 \scriptsize$\uparrow$4.57 &
91.40 \scriptsize$\uparrow$9.10 &
77.65 \scriptsize$\uparrow$6.59 &
96.36 \scriptsize$\uparrow$3.27 &
90.11 \scriptsize$\uparrow$3.56 &
84.70 \scriptsize$\uparrow$13.31 &
87.53 \scriptsize$\uparrow$6.73       \\
 \midrule
AgentPrune                 & 85.07 \scriptsize$\uparrow$4.57      & 91.10 \scriptsize$\uparrow$8.80          & 80.51 \scriptsize$\uparrow$9.45       & 94.65 \scriptsize$\uparrow$1.56               & 90.58 \scriptsize$\uparrow$4.03          & 86.75 \scriptsize$\uparrow$15.36              & 88.09 \scriptsize$\uparrow$7.29         \\
AgentDropout              & 85.62 \scriptsize$\uparrow$5.23        & 91.70 \scriptsize$\uparrow$9.40          & 80.94 \scriptsize$\uparrow$9.88        & 95.60 \scriptsize$\uparrow$2.51               & 91.04 \scriptsize$\uparrow$4.49         & 85.98 \scriptsize$\uparrow$14.59              & 88.48 \scriptsize$\uparrow$7.68         \\
G-designer             & 86.92 \scriptsize$\uparrow$6.53       & 93.80 \scriptsize$\uparrow$11.50         & 81.60 \scriptsize$\uparrow$10.54         & 96.50 \scriptsize$\uparrow$3.41            & 93.10 \scriptsize$\uparrow$6.55          & 88.33 \scriptsize$\uparrow$16.94             & 90.04 \scriptsize$\uparrow$9.24        \\  \midrule
\ourmethod & 
\textbf{89.54 \scriptsize$\uparrow$9.15} & 
\textbf{94.40 \scriptsize$\uparrow$12.10} & 
\textbf{86.45 \scriptsize$\uparrow$15.39} & 
\textbf{98.93 \scriptsize$\uparrow$5.84} & 
\textbf{95.63 \scriptsize$\uparrow$9.08} & 
\textbf{91.74 \scriptsize$\uparrow$20.35} & 
\textbf{92.78 \scriptsize$\uparrow$11.98} \\  
\bottomrule
\end{tabular}
\caption{Performance comparison (\%) on six benchmarks. The best results are highlighted in bold.}\label{tab:performance}
\end{table*}

\section{Experiments}

\subsection{Experimental Setting}

\noindent\textbf{Datasets and Metrics.}
Following~\cite{zhang2024g}, we evaluated \ourmethod on three categories of datasets: \ding{182}~\textbf{General Reasoning:} MMLU~\cite{hendrycks2020measuring};
\ding{183}~\textbf{Mathematical Reasoning:} GSM8K~\cite{cobbe2021training}, MultiArith~\cite{roy2016solving}, SVAMP~\cite{patel2021nlp}, and AQuA~\cite{ling2017program};
\ding{184}~\textbf{Code Generation:} HumanEval~\cite{chen2021evaluating}.
The statistics of the datasets are shown in Table~\ref{tab:dataset_statistics}.

\noindent\textbf{Baselines.}
We compare \ourmethod against various baselines, which can be grouped into four main categories: 
\ding{182}~\textbf{Single-agent methods}, including CoT~\cite{wei2022chain} and Self-Consistency~\cite{wang2022self};
\ding{183}~\textbf{MAS with fixed topologies}, such as Chain, Tree, Complete Graph, and Random Graph~\cite{qian2024scaling};
\ding{184}~\textbf{MAS with Debate} like LLM-Debate~\cite{du2023improving}, where multiple agents iteratively critique and refine responses in a structured process;
\ding{185}~\textbf{MAS with Learnable topologies}, which include AgentPrune~\cite{zhang2024cut}, AgentDropout~\cite{wang2025agentdropout}, and G-Designer~\cite{zhang2024g}.

\noindent\textbf{Implementation Details.}
We access GPT models via the OpenAI API, primarily using \texttt{gpt-4o-2024-08-06} (GPT-4o). We employ a summarizer agent to aggregate the history of dialogue and produce the final solution $a^{(K)}$, with $K=3$ for all baselines across all experiments. The node encoder is implemented using \texttt{all-MiniLM-L6-v2}~\cite{wang2020minilm}, with the embedding dimension set to $D=384$. The hyperparameter $\alpha$ is set to $0.2$ for all experiments. Following classical configurations in LLM-MAS~\cite{zhuge2024gptswarm, yin2023exchange, zhang2024g}, we provide explicit agent profiles for multi-agent methods and use GPT-4 to generate these profile pools. For all datasets, we use $B \in \{40,60\}$ queries for model training.

\subsection{Experimental Results}

\noindent\textbf{Performance Comparison.}
The comparison results are presented in Table~\ref{tab:performance}, from which we have the following observations. \ding{182}~\ourmethod achieves the best performance across all six benchmarks, consistently outperforming a wide range of baselines. The superior performance demonstrates the effectiveness of the autoregressive graph generation paradigm in MAS topology design. \ding{183}~Compared to the strongest learning-based baseline, G-Designer, \ourmethod shows a significant performance gain. For instance, on AQuA, \ourmethod achieves an accuracy of 86.45\%, surpassing G-Designer by a substantial margin of 4.85\%. \ding{184}~When compared to debate-based methods like LLM-Debate, \ourmethod demonstrates a remarkable improvement of 8.8\% on AQuA and 2.66\% on GSM8K. This highlights the inefficiency of the fixed and all-to-all communication protocol. 

\begin{figure}[t]
  \centering
  \subfloat[Token cost of MMLU]{%
    \label{subfig:tokenMMLU}%
    \includegraphics[width=0.5\columnwidth]{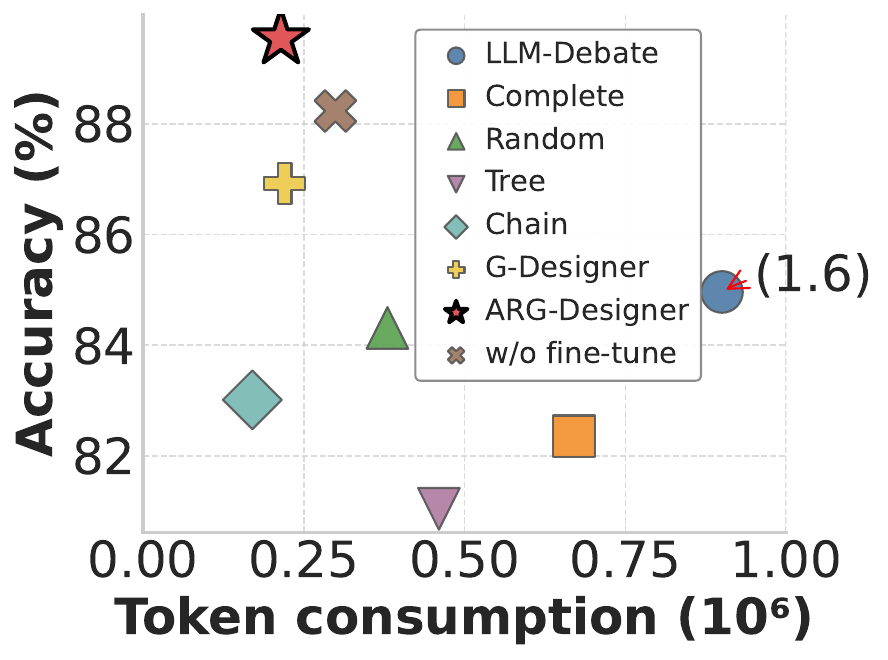}%
  }\hfill
  \subfloat[Token cost of GSM8K]{%
    \label{subfig:tokenGSM8K}%
    \includegraphics[width=0.5\columnwidth]{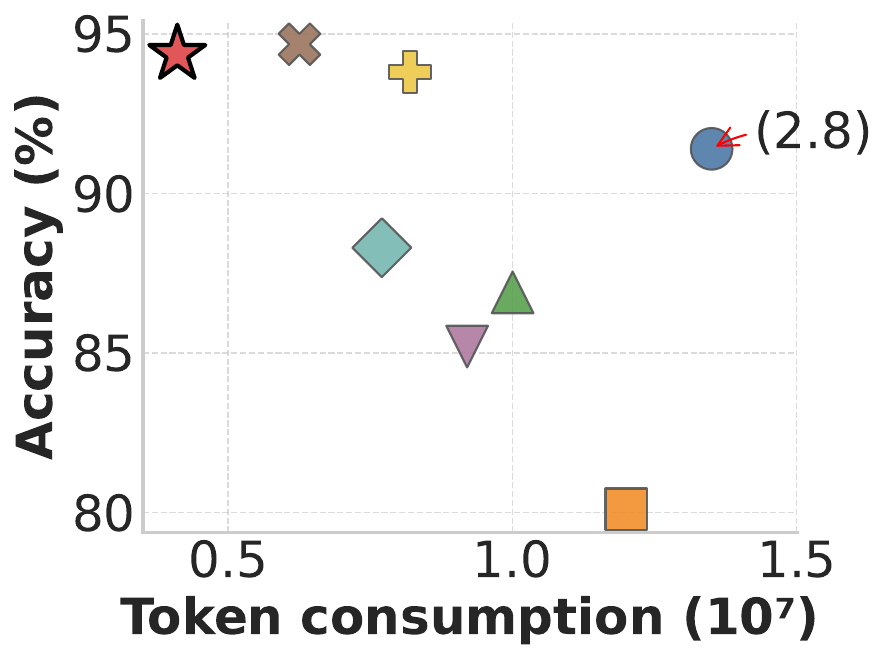}%
  }
  \caption{The prompt token cost comparison.}\label{fig:eff}
\end{figure}

\noindent\textbf{Token Efficiency.}\label{sec:token_efficiency}
A key benefit of \ourmethod is its ability to generate tailored topologies for different tasks, which prevents unnecessary complexity and thus minimizes token consumption. Fig.~\ref{subfig:tokenMMLU} and \ref{subfig:tokenGSM8K} illustrate the trade-off between performance and token cost. We can observe that: \ding{182}~\ourmethod elegantly balances efficiency and performance. On the GSM8K dataset, \ourmethod is the most token-efficient method, using only 4.1e6 tokens, while achieving a top-tier accuracy of 94.40\%. It surpasses the strong G-Designer baseline in accuracy while using approximately 50\% fewer tokens. \ding{183}~A general trend where more complex communication structures, such as the dialogue-based LLM-Debate, achieve relatively good performance but at an extremely high token cost. \ding{184}~A direct comparison between \ourmethod and its \textbf{w/o fine-tune} variant further underscores the value of the efficiency fine-tuning phase. For instance, on MMLU, fine-tuning improves accuracy from 88.23\% to 89.54\% while simultaneously cutting token usage by nearly 30\%. On GSM8K, it reduces token consumption by a massive 34\% (from 6.25e6 to 4.1e6). This demonstrates that our two-phase training strategy is highly effective at optimizing for both performance and efficiency.

\begin{figure*}
    \centering
    \subfloat[Robustness against prompt injection attacks]{
        \includegraphics[height=0.108\textheight]{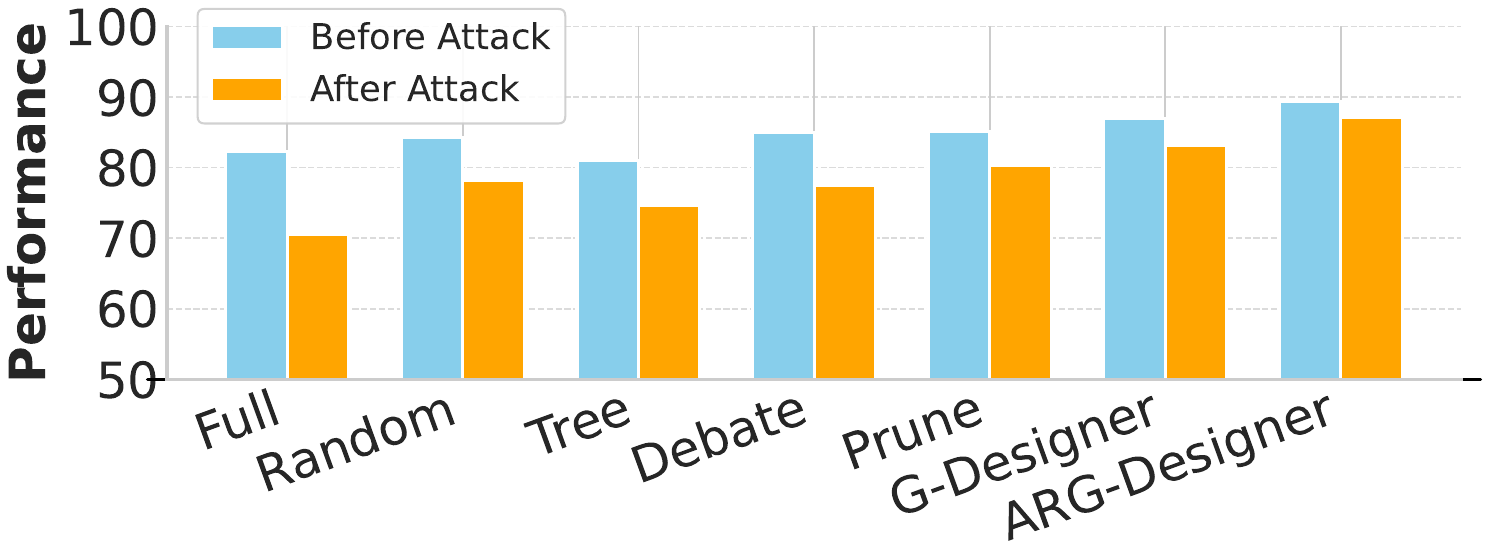}
        \label{subfig:robust}
    }
    \hfill
    \subfloat[Role extensibility on MMLU]{
        \includegraphics[height=0.108\textheight]{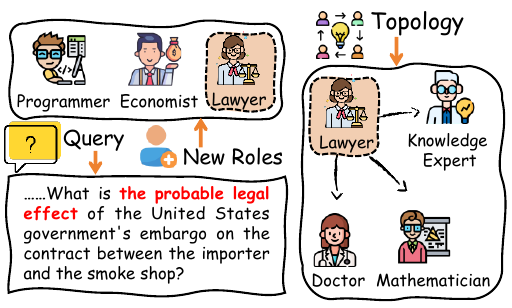}
        \label{subfig:exten}
    }
    \hfill
    \subfloat[Cases of G-Designer and \ourmethod]{
        \includegraphics[height=0.108\textheight]{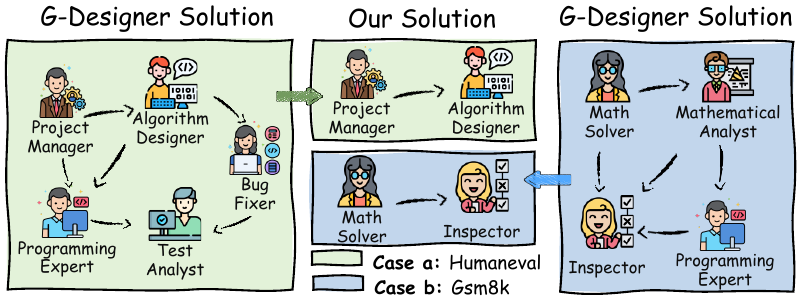}
        \label{subfig:case}
    }
    \caption{The robustness, extensibility of \ourmethod and case studies.}
\end{figure*}

\noindent\textbf{Ablation Study.}
To validate the effectiveness of key components in \ourmethod, we investigate three major variants of our model: \ding{182}~\textbf{w/o fine-tune}, where efficiency fine-tuning phase is removed. \ding{183}~\textbf{w/o task emb.}, where the influence of task embedding $\mathbf{f}_\mathcal{Q}$ is set to be 0. \ding{184}~\textbf{w/o hist. emb.}, where the historical embeddings are removed from the generation model. 
The results in Table~\ref{tab:ablation} demonstrate the contribution of each component: \ding{182}~\ourmethod achieves the highest average score of 91.89, significantly outperforming a vanilla baseline (78.02) that lacks these sophisticated mechanisms. \ding{183}~The absence of task-specific guidance (\textbf{w/o task emb.}) causes the most significant performance degradation, which highlights that conditioning on the task is crucial for generating a bespoke and effective collaboration topology. \ding{184}~The removal of historical embeddings (\textbf{w/o hist. emb.}) also leads to a noticeable decline to 90.64, confirming the value of modeling dependencies between agents. \ding{185}~Interestingly, the performance of the \textbf{w/o fine-tune} variant is highly competitive. This indicates that the model can learn fundamental collaboration patterns from the first stage. Nevertheless, the fine-tuning bring significant efficiency benefits (see Fig.~\ref{fig:eff}), while also slightly improving performance. 

\begin{table}[t]
\centering
\small
\setlength{\tabcolsep}{1pt}{
\begin{tabular}{l|p{1.33cm}<{\centering}p{1.33cm}<{\centering}p{1.65cm}<{\centering}|p{1.33cm}<{\centering}}
\toprule
\textbf{Method} & \textbf{MMLU} & \textbf{GSM8K} & \textbf{HumanEval} & \textbf{Average} \\ 
\midrule
Vanilla  & 80.39 & 82.30 & 71.39 & 78.02 \\
\ourmethod  & \textbf{89.54} & 94.40 & \textbf{91.74} & \textbf{91.89} \\ \midrule
w/o fine-tune & 88.23 & \textbf{94.70} & 90.91 & 91.28 \\
w/o task emb. & 86.93 & 93.10 & 89.26 & 89.76 \\
w/o hist. emb. & 88.23 & 93.60 & 90.08 & 90.64 \\
\bottomrule
\end{tabular}
}
\caption{Results of ablation study.}
\label{tab:ablation}
\end{table}

\noindent\textbf{Robustness Analysis.} 
Following~\citet{zhuge2024gptswarm}, we evaluate the robustness of \ourmethod by simulating a system prompt attack, where an adversarial prompt is injected into a single agent to disrupt its function. As illustrated in Fig.~\ref{subfig:robust}, this attack causes significant performance degradation in MAS with fixed and naive topologies. 
In contrast, \ourmethod shows excellent robustness against attacks with the least performance degradation (2.15\%). This resilience emerges from our training objective, which discourages brittle structures and guides the model to construct fault-tolerant topologies with distributed risk and redundant communication paths.

\noindent\textbf{Extensibility Analysis.} 
Furthermore, we examine the model's extensibility. As depicted in Fig.~\ref{subfig:exten}, we introduce several new roles, to the pre-trained model without any retraining. When presented with a legal question from the MMLU regarding a contract embargo, \ourmethod demonstrates its adaptability. Correctly identifies the high relevance of the newly added ``Lawyer'' role and dynamically generates a collaboration graph placing the lawyer at its core, coordinating with other experts. This case vividly illustrates that \ourmethod can seamlessly scale its capabilities by integrating new knowledge, creating effective specialized team structures on the fly.

\noindent\textbf{Case Study.}
To further illustrate the advantages of \ourmethod over learning-based baselines (e.g., G-Designer~\cite{zhang2024aflow}), we conduct a comparative case study on representative cases in HumanEval and GSM8K. As shown in Fig.~\ref{subfig:case}. The key difference lies in the flexibility of composition versus the static design. G‑Designer requires a predefined and fixed set of agents and a fixed agent count. Its solution graphs remain within this rigid template, regardless of task complexity. \ourmethod, in contrast, dynamically generates both the roles and their communication links from an extensible role pool. It adapts the number of agents and connections based on task needs. Therefore, \ourmethod constructs more efficient collaboration graphs with fewer agents and messages, cutting token usage without sacrificing accuracy.

\section{Related Work}
\subsection{Autoregressive Graph Generation}
Autoregressive models are the cornerstone of graph generation, tackling the intractable joint probability of a graph by factorizing it into a tractable sequence of conditional probabilities for nodes and edges~\cite{you2018graphrnn, liao2019efficient}. 
However, the design of these models is critically dependent on the choice of node ordering. Initial models like GraphRNN~\cite{you2018graphrnn} pioneered this sequential approach but relied on simple, fixed node orderings like Breadth-First Search (BFS). Subsequent work such as GRAN~\cite{liao2019efficient} showed that exploring various ad-hoc ordering schemes could improve performance, but highlighted the theoretical challenge that these orderings can lead to a loose variational bound. To resolve this, GraphGEN~\cite{goyal2020graphgen} proposed using a single canonical order for a graph, although this creates a new mismatch problem, as the generation process itself may not follow this canonical path. Beyond the pivotal node-ordering problem, other research has focused on improving scalability, with methods like BiGG~\cite{dai2020scalable} reducing generation time complexity and others proposing more scalable graph representations~\cite{jang2023simple}. Another key direction is conditional generation, where models like CCGG~\cite{ommi2022ccgg} learn to produce graphs based on a given class label, and hybrid architectures have also been explored, for instance by combining autoregressive models with diffusion models~\cite{kong2023autoregressive}.

\subsection{LLM-based Multi-Agent System}
The advent of powerful Large Language Models (LLMs) has catalyzed a shift from single-agent systems to Multi-Agent Systems (MAS) for tackling complex problems~\cite{qian2024scaling, zhu2024large}. The success of these systems hinges on their collaboration topology, the design of which has thus become a central research problem~\cite{zhuge2024gptswarm, zhang2024aflow}.

Research into topology design has evolved from static to adaptive structures. The initial approaches adopted fixed and manually-designed topologies, such as chains that enforce a sequential workflow~\cite{wei2022chain, hong2023metagpt} or trees that facilitate structured exploration and deliberation~\cite{yao2023tree}. While foundational, the inherent rigidity of these static structures limits their adaptability across diverse tasks, often leading to sub-optimal performance.
To address this, a recent line of work has focused on learning adaptive communication graphs. For instance, some approaches start with a dense, fully-connected graph and learn to prune it in a task-aware manner. AgentPrune~\cite{zhang2024cut} learns to remove redundant communication links, while AgentDropout~\cite{wang2025agentdropout} applies a dynamic dropout technique to both agents and edges. To move beyond simple pruning, more advanced methods leverage the expressive power of graph neural networks (GNNs), which have become a standard for learning on graph-structured data~\cite{li2024noise,chen2025uncertainty}. G-Designer~\cite{zhang2024g}, for instance, employs a GNN-based autoencoder to directly generate a query-dependent communication structure. While these methods achieve task-adaptivity in structuring communication, they are still fundamentally constrained by the initial template. This rigidity prevents the MAS from being truly bespoke, leading to the critical issues of \textit{redundant composition} and \textit{limited extensibility}. 

Drawing inspiration from the real-world practices of recruiting teams incrementally, our work departs fundamentally from modifying a predefined template. We instead propose \textit{\textbf{autoregressive graph generation}}, a paradigm that constructs the collaboration graph from scratch. This allows our method \ourmethod, to jointly determine both the system's composition and structure, dynamically selecting agents from an extensible pool and establishing optimal communication links in a truly task-adaptive manner.

\section{Conclusion}
In this work, we addressed redundant composition and limited extensibility in template graph modification approaches by reformulating collaboration topology design as autoregressive graph generation. We introduced {\ourmethod}, a novel autoregressive model that constructs collaboration graphs from scratch, conditioned on natural language task queries. Our approach dynamically determines agent numbers, selects roles from an extensible pool, and establishes optimal communication links, creating bespoke MAS topologies tailored to specific task demands. Extensive experiments across six benchmarks demonstrate that \ourmethod consistently outperforms existing methods, achieving state-of-the-art performance while maintaining superior token efficiency. 

\newpage
\section*{Acknowledgments}
The work of S. Pan was partially supported by the Australian Research Council (ARC) under Grant Nos. DP240101547 and FT210100097.

\bibliography{aaai2026}

\end{document}